\documentstyle[art12]{article}
\begin{document}
  \begin{center}
{\large\bf Estimating  the contribution from different ionospheric \\
regions to the TEC response to the solar flares using data \\
from the international GPS network \\ }
L.A. Leonovich, E. L. Afraimovich, E.B. Romanova and A.V. Taschilin \\
Institute of Solar-Terrestrial Physics SD RAS, Irkutsk, Russia \\
  \end{center}
\date{}

\begin{abstract}
This paper proposes a new method for estimating the contribution
from different ionospheric regions to the response of total
electron content variations to the solar flare, based on data from
the international network of two-frequency multichannel receivers
of the navigation GPS system. The method uses the effect of
partial "shadowing" of the atmosphere by the terrestrial globe.
The study of the solar flare influence on the atmosphere uses GPS
stations located near the boundary of the shadow on the ground in
the nightside hemisphere. The beams between the satellite-borne
transmitter and the receiver on the ground for these stations pass
partially through the atmosphere lying in the region of total
shadow and partially through the illuminated atmosphere. The
analysis of the ionospheric effect of a powerful solar flare of
class X5.7/3B that was recorded on July 14, 2000 (10:24 UT,
N22W07) in quiet geomagnetic conditions (Dst = -10 nT) has shown
that about 20\% of the TEC increase correspond to the ionospheric
region lying below 100 km, about 5\% refer to the ionospheric
E-region (100-140 km), about 30\% correspond to the ionospheric
F1-region (140-200 km), and about 30\% to regions lying above 300~km.
\end{abstract}
\section*{Keywords}
Solar Flare, GPS, ionosphere

\section{Introduction}

The enhancement of X-ray and ultraviolet (UV) emission that is
observed during chromospheric flares on the Sun immediately causes
an increase in electron density in the ionosphere. These density
variations are different for different altitudes and are called
Sudden Ionospheric Disturbances, SID (Davies,1990; Donnelly, 1969). SIDs
are generally recorded as the short wave fadeout, SWF
(Stonehocker, 1970), sudden phase anomaly, SPA (Ohshio, 1971), sudden
frequency deviation, SFD (Donnelly, 1971; Liu et al., 1996),
 sudden cosmic noise
absorption, SCNA (Deshpande and Mitra, 1972), sudden enhancement/decrease of
atmospherics, SES (Sao et al., 1970). Much research is devoted to SID
studies, among them a number of thorough reviews (Mitra, 1974; Davies, 1990).

Highly informative technique is the method of Incoherent Scatter
(IS). The Millstone Hill IS facility recorded a powerful flare on
August 7, 1972 (Mendillo and Evans, 1974a). The measurements were made in the
height range from 125 to 1200~km. The increase of local electron
density $N_{e}$ made up 100\% at 125~km altitude and 60\% at 200~
km.

Using the IS method Thome and Wagner (1971) obtained important evidence of
the height distribution of the increase in $N_{e}$ at the time of
the May 21 and 23, 1967 flares. A significant increase of $N_{e}$
was recorded in the E-region, up to 200\%, which gradually
decreases in the F-region with the increasing height, down to
10-30\%, and remains distinguishable up to 300 km. The earliest
increase of $N_{e}$ begins in the E-region, and at higher
altitudes it is observed with a delay which is particularly
pronounced at F-region heights.

A sudden increase in total electron content (TEC) can be measured
using continuously operating radio beacons installed on
geostationary satellites. On August 7, 1972, Mendillo et al. (1974b) were
the first to make an attempt to carry out global observations of
the solar flare using 17 stations in North America, Europe, and
Africa. The observations covered an area, the boundaries of which
were separated by $70^{\circ}$ in latitude and by 10 hours in
local time. For different stations,
the absolute value of the TEC  increase $\Delta I$
varies from $ 1.8 \cdot 10^{16}$ to
$ 8.6 \cdot 10^{16} \mbox{el} \cdot \mbox{m}^{-2}$,
which corresponds to 15-30\% of the TEC.
Investigations revealed a latitudinal dependence of the TEC increase value.
At low latitudes, it was higher compared with high latitudes.
Besides, the authors point out the absence of a connection between the
TEC increase value and the solar zenith angle.

The advent and evolution of a Global Positioning System (GPS) and
also the creation on its basis of widely branched networks of GPS
stations (at least 900 sites at the August of 2001, the data from
which are placed on the Internet) opened up a new era in remote
ionospheric sensing. High-precision measurements of the TEC along
the line-of-sight (LOS) between  the receiver on the ground and
transmitters on the GPS system satellites covering the reception
zone are made using two-frequency multichannel receivers of the
GPS system at almost any point of the globe and at any time
simultaneously at two coherently coupled frequencies
$f_{1}=1575.42$ MHz and $f_{2}=1227.60$ MHz.

The sensitivity of phase measurements in the GPS system is
sufficient for detecting irregularities with an amplitude of up to
$10^{3}-10^{4}$ of the diurnal TEC variation. This makes it
possible to formulate the problem of detecting ionospheric
disturbances from different sources of artificial and natural
origins. The TEC unit (TECU) which is equal to $10^{16}$
$\mbox{el} \cdot \mbox{m}^{-2}$ and is commonly accepted in the
literature, will be used throughout the text.

Afraimovich (2000a); Afraimovich et al. (2000b, 2001a,b)
 developed a novel
technology of a global detection of ionospheric effects from solar
flares and presented data from first GPS measurements of global
response of the ionosphere to powerful impulsive flares of July
29, 1999, and December 28, 1999. Authors found that fluctuations
of TEC are coherent for all stations on the dayside of the Earth.
The time profile of TEC responses is similar to the time behavior
of hard X-ray emission variations during flares in the energy
range 25-35 keV if the relaxation time of electron density
disturbances in the ionosphere of order 50-100 s is introduced. No
such effect on the nightside of the Earth has been detected yet.

Afraimovich et al. (2001c) and Leonovich et al. (2001) suggested 
a technique for
estimating the ionospheric response to weak solar flares (of X-ray
class C). They obtained a dependence of the ionospheric TEC
increase amplitude (during the solar flare) on the flare location
on the Sun (on the central meridian distance, CMD). For flares
lying nearer to the disk center (CMD $ < 40^{\circ}$), an
empirical dependence of the ionospheric TEC increase amplitude on
the peak power of solar flares in the X-ray range was obtained
(using data from the geostationary GOES-10 satellite).

 This paper is a logical continuation of the series of our publications
(Afraimovich, 2000a; Afraimovich et al., 2000b, 2001a,b,c, 
Leonovich et al., 2001) devoted to the
study of ionospheric effects of solar flares, based on data from
the international GPS network.

A limitation of the GPS method is that its results have an
integral character, as a consequence of which from measurements at
a single site it is impossible to determine which ionospheric
region makes the main contribution to the TEC variation. The
objective of this study is to develop a method which would help
overcome (at least partially) this problem.

\section{Method of determining the shadow altitude $h_{0}$ over the ground}

The method uses the effect of partial "shadowing" of the
atmosphere by the terrestrial globe. Direct beams of solar
ionizing radiation from the flare do not penetrate the region of
the Earths total shadow. GPS stations located near the shadow
boundary on the ground in the nightside hemisphere are used to
investigate the solar flare influence on the ionosphere.  The LOS
for these stations pass partially through the atmosphere lying in
the total shadow region, and partially through the illuminated
atmosphere. The altitude over the ground at which the LOS
intersects the boundary of the total shadow cone, will be referred
to as the shadow altitude $h_{0}$.

Fig. 1 schematically represents the formation of the cone of the
Earth's total shadow (not to scale) in the geocentric
solar-ecliptic coordinate system (GSE): the axis Z is directed to
a north perpendicularly planes of an ecliptic, the axis X - on the
Sun, the axis Y is directed perpendiculary to these axes. For
definition of the shadow altitude $h_{0}$ it is necessary to know
coordinates of a cross point C of the  LOS and the shadow
boundary.

The primary data are the geographical coordinates of station GPS
on the Earth (Fig.1; a point P): an elevation angle and azimuth of
LOS on a satellite GPS, toward the north clockwise, for the time
(UT) corresponding to the phase of solar flare maximum in the
X-ray range. These coordinates are converted to the  Cartesian
coordinate system where the Cartesian coordinates of the GPS
station on the ground and the coordinates of the subionospheric
point (at 300 km altitude) are calculated. Next, we use the
geocentric solar-ecliptic coordinate system following the
technique reported by Sergeev and Tsyganenko (1980). 
To determine the coordinates
of the point C we solve a system of equations: the equation of
cone (of total shadow), and the equation of a straight line (LOS)
specified parametrically. After that, from the resulting point C
we drop a perpendicular to the ground and calculate its length
(Fig. 1, line $h_{0}$). The value of $h_{0}$, thus obtained, is
just the shadow altitude.

\section{ Method  of determining the TEC increase in the ionosphere using
data from the global GPS net work}

This paper exemplifies an analysis of the ionospheric effect of a
powerful solar flare of class X5.7/3B recorded on July 14, 2000
(10:24 UT, N22W07) under quiet geomagnetic conditions (Dst = -10
nT). The time profile of soft X-ray emission in the range 1-8 \AA
~(GOES-10 data) at the time of the flare is presented in Fig. 2a.

To determine the TEC increase in the ionosphere we used the data
from the international GPS network. The GPS technology provides a
means of estimating the TEC variations $I_{0}(t)$ on the basis of
TEC  phase measurements made with each of the spatially separated
two-frequency GPS receivers using the formula (Calais and Minster, 1996):

\begin{equation}
I_{0}(t)=\frac{1}{40{.}308}\frac{f^2_1f^2_2}{f^2_1-f^2_2}
                          [(L_1\lambda_1-L_2\lambda_2)+const+nL],
\end{equation}

where $L_1\lambda_1$ and $L_2\lambda_2$  are the increments of the
radio signal phase path caused by the phase delay in the
ionosphere (m); $L_1$ and $L_2$ stand for the number of complete
phase rotations, and $\lambda_1$ and $\lambda_2$ are the
wavelengths (m) for the frequencies $f_{1}$ and $f_{2}$,
respectively; $const$ is some unknown initial phase path (m); and
$nL$ is the error in determining the phase path (m).

Input  data used in the analysis include series of the oblique
value of TEC $I_{0}(t)$, as well as corresponding series of
elevations $\theta$ and azimuths of LOS to the satellite. These
parameters are calculated using our developed CONVTEC program to
convert standard (for the GPS system) RINEX-files received via the
Internet. Input series of TEC $I_{0}(t)$ are converted to the
vertical value following a well-known technique (Klobuchar, 1986).

\begin{equation}
 I(t) = I_{0} \cdot cos \left[arcsin\left(\frac{R_E}{R_E
+ h_{ma x}}cos\theta\right)\right]
\end{equation}

where $R_{E}$ is Earth's radius; and $h_{max}$ is the height of
the ionospheric $F2$-layer maximum.

Variations of the regular ionosphere, and also trends introduced
by  the motion of the satellite are eliminated using the procedure
of removing the trend defined as a  polynomial of the third order
on a given temporary interval.

Figs. 2b and 2d present the typical time dependencies of the
vertical TEC $I(t)$ for sites GPS WDLM (PRN02, shadow altitude
$h_{0} = 17$ km) and LEEP (PRN07, shadow altitude $h_{0}= 586$
km). Time dependencies of the TEC $\Delta I(t)$ response, with the
trend removed for these series, are presented in Figs. 2c and 2e,
respectively.

\section{  Results and discussion }

The TEC response  to the solar flare was analyzed for 45 GPS
stations. Detailed information about the GPS stations and analysis
results is summarized in Table 1: names of GPS receiving stations
(Site), number of the GPS satellite from which the signal is
received (PRN), shadow altitude above the ground ($h_{0}$),
absolute increase of TEC $\Delta I$, relative increase of TEC
($\Delta I(t)/ \Delta I_{00}(t)$, and geographical coordinates of
GPS stations (latitude, longitude). The increase of TEC $\Delta
I_{00}$ corresponds to the amplitude of the TEC increase measured
at the station lying at the shadow boundary on the ground
($h_{0}=0$).

Fig. 3b illustrates  examples of time dependencies of the TEC
$\Delta I_{0}$ response for LOS to the satellite which intersect
the boundary of the shadow cone at different heights $h_{0}$
during the solar flare of July 14, 2000. Fig. 3b (left) presents
the values of these altitudes, and (right) the names of
corresponding stations. For a better visualization, the
dependencies are drawn by lines of a different thickness. It
should be noted that the response remains pronounced when the
shadow altitude  exceeds significantly the electron density peak
height in the ionosphere. For station GUAM (PRN26, height of the
shadow boundary $h_{0}= 885$ km), the response amplitude exceeds
the background oscillation amplitude by more than a factor of 2.

It is evident from  Fig. 3 that the wave phase (time of the
response maximum) is different at different altitudes $h_{0}$. On
the one hand, this phenomenon can be caused by the interference of
the response with background fluctuations; on the other, this can
be due to the fact that at different heights different wavelengths
of ionizing radiation are observed, which, in turn, can have
independent time characteristics.

The dependence   of the absolute TEC increase on the altitude
$h_{0}$ for all the cases under consideration is plotted in Fig.
4a. The dependence of the relative TEC increase $\Delta I(t)/
\Delta I_{00}$ on the altitude $h_{0}$ during the solar flare is
shown as a percentage in Fig. 4b. The TEC increase $\Delta
I_{00}(t)$ corresponds to the amplitude of the TEC increase
measured at the station lying at the shadow boundary on the ground
($h_{0}=0$).

Fig. 4b suggests that about 20\% of the TEC increase correspond to
the ionospheric region lying below 100 km, about 5\% refer to the
ionospheric E-region (100-140 km), about 30\% correspond to the
ionospheric F1-region (140-200 km), and about 30\% to regions
lying above 300 km. We found that a rather significant
contribution to the TEC increase is made by ionospheric regions
lying above 300 km.

The estimate obtained is consistent with the findings reported by
Mendillo and Evans (1974a); Mendillo et al. (1974b). 
The authors of the cited references, based
on investigating the electron density profile in the height range
from 125 km to 1200 km using the IS method, concluded that about
40\% of the TEC increase during the powerful flare on August 7,
1972,  correspond to ionospheric regions lying above 300 km.
However, Thome and Wagner (1971), who used the IS method to investigate the
ionospheric effects from two others powerful solar flares, pointed
out that an increase in electron density associated with the solar
flare was observable to 300 km altitude only. This difference can
be explained by the fact that each particular solar flare is a
unique event which is characterized by its own spectrum and
dynamics in the flare process.

\vspace{0.5cm}      
{\parindent=0cm
{\it Acknowledgements.}
Authors are grateful to V.G. Mikhalkovsky for his assistance in
preparing the English version of the manuscript. This work was
done with support from both the Russian foundation for Basic
Research (grant 00-05-72026) and RFBR grant of leading scientific
schools of the Russian Federation No. 00-15-98509.           % }

\vspace{0.5cm}      
{\bf References}

\vspace{0.5cm}      
 {\bf Afraimovich, E. L.,} GPS global detection of the ionospheric
response to solar flares, {\em Radio Sci.,}  {\bf35,} 1417--1424,
2000a.

 {\bf Afraimovich, E. L.,} Kosogorov,  E. A., and  L. A. Leonovich, The
use of the international GPS network as the global detector
(GLOBDET) simultaneously observing sudden ionospheric
disturbances, {\em Earth Planet. Space,}  {\bf 52,} 1077--1082,
2000b.

 {\bf Afraimovich, E.L.,}  Altyntsev, A.T., Kosogorov, E.A., Larina,
N.S., and  L. A. Leonovich,  Detecting of the Ionospheric effects
of the solar flares as deduced from global GPS network data,
{\em Geomagnetism and Aeronomy,}  {\bf 41,}  208--214, 2001a. %N2

 {\bf Afraimovich, E.L.,}   Altyntsev, A.T., Kosogorov, E.A., Larina,
N.S., and   L. A. Leonovich,  Ionospheric effects of the solar
flares of September 23, 1998 and July 29, 1999 as deduced from
global GPS network data, {\em J. Atm. Solar-Terr. Phys.,}  {\bf 63,}   % N17,
1841-1849, 2001b.

{\bf Afraimovich, E.L.,} Altyntsev, A.T., Grechnev, V.V., and
 L.~A.~Leonovich, Ionospheric effects of the solar flares as
deduced from global GPS network data, {\em Adv. Space Res.,}
 {\bf 27,} 1333-1338, 2001c.         %N6-7,

 {\bf Calais,~E., and J.~B.~Minster,} GPS detection of ionospheric
perturbations following a Space Shuttle ascent, {\em Geophys. Res.
Lett.,}  {\bf 23,} 1897--1900, 1996.

 {\bf Davies~K.,} {\em Ionospheric radio,} Peter Peregrinus, London, %p.580,
1990.

 {\bf Deshpande, S. D. and A.~P.~Mitra, }  Ionospheric  effects of solar
flares,  IV,  electron density profiles deduced from measurements
of SCNA's and VLF phase and amplitude, {\em J. Atmos. Terr.
Phys.,} {\bf 34,} 255--259, 1972.

 {\bf Donnelly, R. F.,} Contribution of X-ray  and  EUV bursts of solar
flares to Sudden frequency deviations, {\em J. Geophys. Res.,}
{\bf 74,} 1873 -- 1877, 1969.

 {\bf Donnelly, R. F.,}   Extreme
ultraviolet  flashes  of  solar  flares  observed via sudden
frequency deviations:  experimental results, {\em Solar Phys.,}
{\bf 20,} 188--203, 1971.

 {\bf Klobuchar,~J.~A.,} Ionospheric time-delay algorithm for
single-frequency GPS users, {\em IEEE Transactions on Aerospace
and Electronics System, AES }, {\bf 23(3)}, 325--331, 1986.   %?------

 {\bf Leonovich, L.A.,}  Altynsev, A.T., Afraimovich, E. L., and
 V.~V.~Grechnev, Ionospheric effects of the solar flares as
deduced from global GPS network data. LANL e-print archive, 2001,
http://arXiv.org/abs/physics/011006.

 {\bf Liu,~J.~Y.,}  Chiu, C.S.,  and C. H. Lin, The solar flare
radiation responsible  for  sudden frequency deviation  and
geomagnetic fluctuation,  {\em J. Geophys.  Res.,}  {\bf 101,}
10855--10862, 1996.

 {\bf Mendillo, M., and J.V. Evans,} Incoherent scatter observations of
the ionospheric response to a large solar flare, {\em Radio Sci.,}
 {\bf 9,} 197-203, 1974a.

 {\bf Mendillo, M.,}  Klobuchar, J.  A.,  Fritz, R.  B.,  da Rosa, A.V.,
Kersley, ~L., Yeh,~K.~C., Flaherty, B.~J.,  Rangaswamy, S.,
Schmid, P. E.,  Evans, J.~V.,  Schodel,  ~J.~P., Matsoukas,~D.~A.,
Koster,~J. ~R., Webster,~A.~R., and P.~Chin, Behavior of the
Ionospheric F Region During the Great Solar Flare of August  7,
1972,  {\em J. Geophys. Res.,}  {\bf 79,} 665--672, 1974b.

 {\bf Mitra,~A.~P.,} {\em Ionospheric effects
of solar flares,} D.Reidel, Norwell, Mass.,    %p.249,
1974.

 {\bf Ohshio, M.,} Negative sudden phase anomaly, {\em Nature,}   {\bf 229,} 239--244,
1971.

 {\bf Sao, K.,} Yamashita, M., Tanahashi, S., Jindoh,  H.,  and
K.~Ohta, Sudden enhancements (SEA) and decreases (DSA) of
atmospherics, {\em J. Atmos. Terr. Phys.,}  {\bf 32,} 1567--1573,
1970.

 {\bf Sergeev, V. A., and N. A. Tsyganenko, } {\em The earth's magnetosphere.
Results of researches on the international geophysical projects.}
'Nauka', Moscow,  (in Russian), 1980.

 {\bf  Stonehocker, G. H.,} Advanced
telecommunication forecasting  technique  in  AGY,  5th.,
Ionospheric forecasting, AGARD Conf. Proc.,  {\bf 29,} 27-31, 1970.  %   ?

 {\bf Thome, G.D, and L. S. Wagner, } Electron density enhancements in the
E and F regions of the ionosphere during solar flares, {\em
J.Geophys. Res.,}  {\bf 76,} 6883--6895, 1971.
}
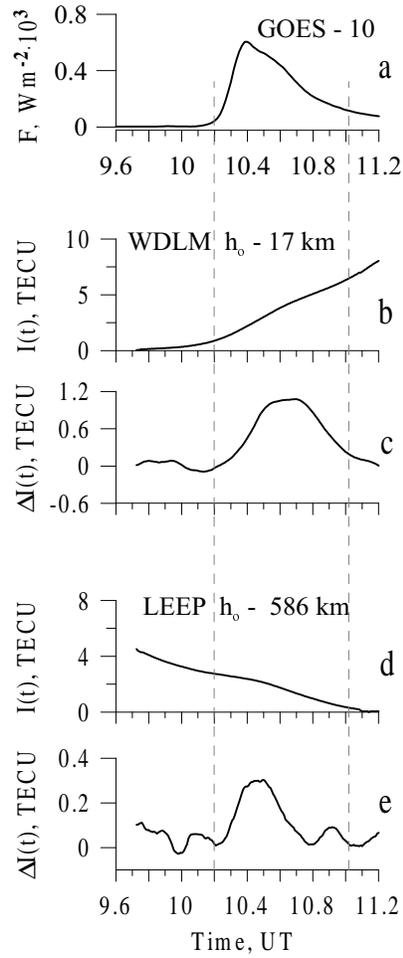
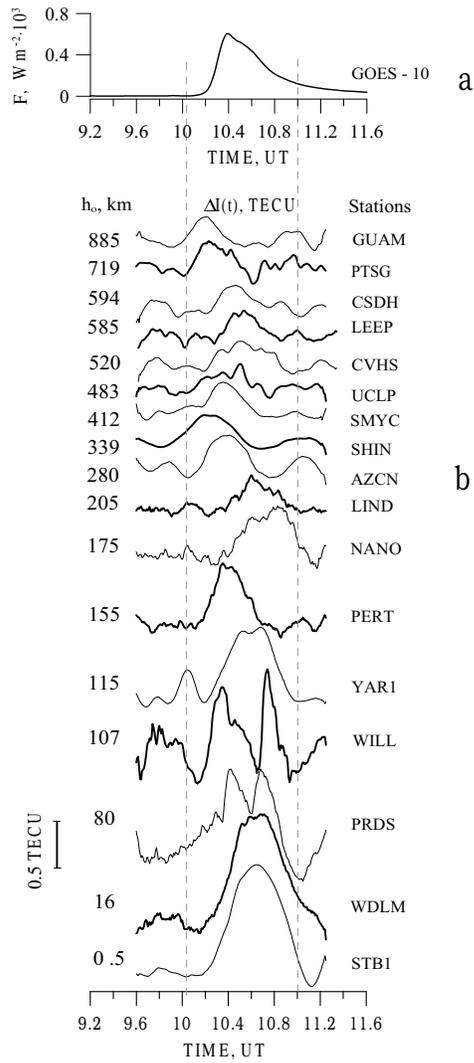
\end{document}